\documentclass[%
reprint, superscriptaddress,
showpacs,preprintnumbers,
dvipsnames,svgnames,table,
 amsmath,amssymb,
 aps]{revtex4-1}
\usepackage{amsmath}
\usepackage{amssymb}
\usepackage{graphicx}
\usepackage[caption=false]{subfig}
\usepackage{float}
\usepackage[left=25mm,right=25mm,top=25mm,bottom=25mm]{geometry}
\usepackage[usenames,dvipsnames]{xcolor}
\usepackage{enumitem}
\usepackage{epstopdf}
\usepackage{hyperref}
\hypersetup{
    colorlinks=true,
    linkcolor=blue,
    }
\usepackage[authormarkup=none]{changes}
\definechangesauthor[name={Stefan}, color=purple]{n}
\definechangesauthor[name={Stella}, color=blue]{s}
\definechangesauthor[name={Valerio}, color=red]{v}
\definechangesauthor[name={Angeline}, color=orange]{a}
\definechangesauthor[name={Caiyu}, color=green]{c}

\newcommand{\ket}[1]{\left|#1\right>}
\newcommand{\bra}[1]{\left<#1\right|}

\newcommand{\pic}[2]{\includegraphics[#1\textwidth]{#2}}
\newcommand{\tr}{\mathrm{tr}}

\newcommand{\ba}{\begin{eqnarray}}\newcommand{\ea}{\end{eqnarray}}

\begin{document}

\title{Violation of all the second laws of thermal operations by inhomogeneous reservoirs}

\author{Angeline Shu}
 \affiliation{Department of Physics, National University of Singapore, 2 Science Drive 3, Singapore 117542.}
 \affiliation{Centre for Quantum Technologies, National University of Singapore, 3 Science Drive 2, Singapore 117543.}
\author{Yu Cai}
 \affiliation{Centre for Quantum Technologies, National University of Singapore, 3 Science Drive 2, Singapore 117543.}
\author{Stella Seah}
 \affiliation{Department of Physics, National University of Singapore, 2 Science Drive 3, Singapore 117542.}
\author{Stefan Nimmrichter}
 \affiliation{Centre for Quantum Technologies, National University of Singapore, 3 Science Drive 2, Singapore 117543.}
\author{Valerio Scarani}
 \affiliation{Department of Physics, National University of Singapore, 2 Science Drive 3, Singapore 117542.}
 \affiliation{Centre for Quantum Technologies, National University of Singapore, 3 Science Drive 2, Singapore 117543.}
\begin{abstract}
In the resource theory of thermodynamics, the decrease of the free energy based on von Neumann entropy is not a sufficient condition to determine free evolution. Rather, a whole family of generalised free energies $F_{\alpha}$ must be monotonically decreasing. We study the resilience of this result to relaxations of the framework. We use a toy collisional model, in which the deviations from the ideal situation can be described as arising from inhomogeneities of local fields or temperatures. For any small amount of perturbation, we find that there exist initial states such that both single-shot and averaged values of $F_{\alpha}$ do not decrease monotonically for all $\alpha>0$. A geometric representation accounts for the observed behavior in a graphic way. 
\end{abstract}
\maketitle

\section{Introduction}

Foundationally, thermodynamics is a theory of states and their transformations. In quantum information science, the same can be said for entanglement theory. This analogy was noticed very early, and has later resulted in the development of the broad framework of resource theories. The \textit{resource theory of thermal operations} is a formalisation of the thermodynamics of systems in contact with thermal baths \cite{Brandao_2013}. The basic framework having been explored in depth (see \cite{Gour_2015,Goold_2016} for reviews), recent works have been focusing on the resilience of the results when some of the assumptions are relaxed \cite{Sparaciari_2017,Meer_2017,Mueller_2017,Baumer_2017}. Our work contributes to this line of research. 

With two thermal baths at different temperatures, one can run an engine: therefore, the lack of resources is described by what can be achieved with a single thermal bath at temperature $T$. More precisely, the free states are the thermal states $\tau$ at temperature $T$ and the free operations $U$ are those that conserve the total energy. Both notions are defined with respect to a reference Hamiltonian, usually taken as $H=H_S+H_R$ where $S$ indicate the system and $R$ a reservoir of auxiliary systems. Then, thermal states read $\tau=\tau_S\otimes\tau_R$ where $\tau_X=e^{-\beta H_X}/Z_X$, $Z_X=\tr(e^{-\beta H_X})$ and $\beta=1/kT$. Free operations $U$ are required to satisfy \footnote{We notice that, in a dynamical system, the approach to energy conservation is different. There, the whole evolution is generated by the Hamiltonian that includes interactions. Energy is then conserved if the total Hamiltonian is time-independent; if it is time-dependent, there is no reason for energy to be conserved.} \ba\label{conserve}
[H,U]=0\,.
\ea
If the system is prepared in an arbitrary state $\rho$, a \textit{free evolution} (i.e. one that can be achieved without further resources) is then of the form
\ba\label{dynamics}
\mathcal{E}[\rho]&=& \mathrm{tr}_{R}\bigl[U (\rho\otimes\tau_{R})U^{\dagger}\bigr]\,.
\ea
The set of criteria under which a free evolution is possible can be seen as the analog of the second law of thermodynamics. So far, it has not been possible to reduce these criteria to a single one \cite{Gour_2015}. Brandao and coworkers \cite{Brandao_2015} based the \textit{second laws of thermal operations} on the monotonical decrease of a continuous family of \textit{generalised free energies} $F_{\alpha}(\rho||\tau_{S})$, $\alpha\in\mathbb{R}$. For states $\rho=\sum_jp_j\ket{j}\bra{j}$ diagonal in the eigenbasis of $H_S$, the expression is
\ba F_{\alpha}(\rho||\tau_{S})&=&k_{\rm B}T\, \left[D_{\alpha}(\rho||\tau_{S})- \log Z_S\right]\,.\ea where the R\'enyi divergence is given by
\begin{align} 
D_{\alpha}(\rho||\tau_{S})=\frac{\text{sgn}(\alpha)}{\alpha-1}\log\sum_{j}p_{j}^{\alpha}q_{j}^{1-\alpha}\label{eq:RenyiDiv}
\end{align} with $q_j=e^{-\beta E_j}/Z_S$ the eigenvalues of $\tau_S$.


These laws having been established, their robustness under modifications of the framework has been the object of recent studies. Imperfections in the initial and target state can be accounted for by using a suitable family of smoothed generalized free energies \cite{Meer_2017}. If the evolution involves a catalyst, whose state is allowed to be correlated with that of the system at the end of the evolution, then only the monotonicity of $F_{1}$ survives as both a necessary and sufficient condition for the state transitions \cite{Mueller_2017}. 

In this paper, we explore the stability of these second laws under deviations from the exact conservation constraint \eqref{conserve}. We use a specific toy model of a thermalising channel, initial states that are diagonal in the eigenbasis of the Hamiltonian, and what is arguably the simplest form of the perturbation. Even in this restricted context, we find that, however small the perturbation, there exist initial states whose evolution violates \textit{all} the second laws except for $\alpha=0$. The violation is not only visible in single-shot realisations, but also in the average over several such realisations.

The plan of the paper is as follows. In section \ref{sec:ToyMod}, we introduce the toy model and derive analytical results for the average over several realisations. Section \ref{sec:results} presents and discusses the results obtained for qubit (\ref{sec:qubit}) and qutrit systems (\ref{sec:qutrit}). The physical relevance of the model is discussed in Section \ref{sec:PhyRel} before concluding in Section \ref{sec:conclusion}.


\section{Toy model of the dynamics}
\label{sec:ToyMod}

\subsection{Structure of the model}

We consider a collisional model that defines a thermalising channel \cite{Valerio_2002}. The system is a qudit, and the bath consists of qudits labelled by $r\in\{1,2,...,N\}$ with $N\gg 1$. The Hamiltonian is given by
\ba
H=H_S+H_R(\underline{g})&=&g_0 s_z^{(S)}\,+\,\sum_{r=1}^N g_r s_z^{(r)}
\ea where $g_0>0$, $\underline{g}=(g_1,...,g_r)$ and where $s_z$ is the operator representing the spin in the direction $z$. For every qudit, the eigenstates of $s_z$ for the eigenvalue $\hbar \big(j-\frac{d-1}{2}\big)$ is denoted by $\ket{j}$ with $j\in\{0,1,...,d-1\}$.


The qudits of the reservoir are prepared in the thermal state $\tau_R=\bigotimes_r \tau_r$ with $\tau_r=e^{-\beta g_r s_z}/Z_r$. The collisional character of the model is seen in the interaction, that is taken to be of the form $
U=U_{S,N}U_{S,N-1}...U_{S,1}$. We assume that all two-body interactions $U_{S,r}$ are given by the partial swap with mixing angle $\theta$: with $\ket{jj'}\equiv\ket{j}_{S}\ket{j'}_{r}$,
\ba
U_{S,r}&=&\sum_{j=1}^d \Big[\ket{jj}\bra{jj} \nonumber\\
&&+\sum_{j'\neq j} \big(\cos\theta\ket{jj'}+i \sin\theta\ket{j'j}\big)\bra{jj'}\Big]\,. \label{partialswap}
\ea

\subsection{Free thermal operation and perturbation}

The unperturbed model is the one where $g_r=g_0$ for all $r$. In this case, $[H,U]=0$, since $U$ couples only degenerate eigenstates of $H$. Thus, this defines a free thermal operation. For any initial state, the dynamics \eqref{dynamics} can be solved analytically:
\begin{align}
    \rho_{r}&=\rho_{r-1} \cos^2 \theta + \tau_S \sin^2 \theta \nonumber \\
    &=\tau_S - \left(\tau_S-\rho_{0}\right)\cos^{2r} \theta. \label{eq:Analytic}
    \end{align}
In particular, the state of the system converges to the thermal state $\tau_S$ in the limit $N\rightarrow\infty$. Recently, B\"{a}umer et al.~\cite{Baumer_2017} built on this model to derive an extension of the optimal protocol for work extraction \cite{Skrzypczyk_2014}. In their scheme, there are additional storage qudits. Even if the unitary is imperfect and thermalisation is partial, the maximal amount of energy $\Delta F_{1}=F_{1}(\rho||\tau_{S})-F_{1}(\tau_{S}||\tau_{S})$ can be extracted with sufficiently many steps.

In order to study deviations from \eqref{conserve}, we consider the scalar perturbation
\ba
g_r\equiv g(\delta_r)&=&g_0\,(1+\delta_r)
\ea where each $\delta_r$ is a random number drawn from a Gaussian distribution $G(\delta)$ centered at $\overline{\delta} = 0$ with variance $\overline{\delta^2}=\Delta^2$. The physical relevance of this perturbation is discussed in section \ref{sec:PhyRel}. The eigenstates of $H$ remain the same as in the unperturbed case; however, as soon as $\delta_r\neq 0$, $\ket{j}_{S}\ket{j'}_{r}$ and $\ket{j'}_{S}\ket{j}_{r}$ are no longer degenerate. Then, $[H,U]\neq 0$ as desired. Nevertheless $\mathrm{tr}(HU\rho\otimes\tau_{R}U^{\dagger})\approx\mathrm{tr}(H\rho\otimes\tau_R))$ because the fluctuations will cancel on average.

The collisions are now described by a stochastic map which yields individual random trajectories for the system with no fixed point. There is no closed analytical formula of the dynamics for a specific realisation. However, for the ensemble-averaged states
\ba\label{ensavg}
\overline{\rho}_r&=&\int_{-\infty}^{\infty}G(\delta)\rho_r(\delta)\,\mathrm{d}\delta
\ea the dynamics is
\begin{align}
    \overline{\rho}_{r}&= \overline{\tau}- \left(\overline{\tau}-\rho_{0}\right) \cos^{2r} \theta\label{eq:Analytic_Noise}
\end{align}
where the ensemble-averaged thermal state $\overline{\tau}$ is obtained by replacing $\rho_r(\delta)$ with $\tau(\delta)=e^{-\beta g_0(1+\delta)s_z}/\tr(e^{-\beta g_0(1+\delta)s_z})$ in \eqref{ensavg}. Since $\tau(\delta)$ is not a linear function of $\delta$, $\overline{\tau}$ is different from $\tau_S$, and the latter appears in the expressions of the free energies. This is going to be a key observation for the interpretation of our results.



\section{Results and discussion}
\label{sec:results}

In what follows, we shall particularize our study to qubits, then to qutrits. We shall study $D_{\alpha}(\rho||\tau_{S})$, omitting the energy factor $k_{\rm B}T$ and the constant offset $\log{Z_{S}}$. For every step $r$, we shall focus on three quantities:
\begin{itemize}
    \item The single-shot values ${D}_{\alpha}(\rho_{r}||\tau_{S})$ for some specific realisation of the $\underline{\delta}$, obtained numerically
    \item The values  $D_{\alpha}(\overline{\rho}_{r}||\tau_{S})$ for the ensemble-averaged states, obtained from \eqref{eq:Analytic_Noise}.
    \item The ensemble averages $\overline{D}_{\alpha}(\rho_{r}||\tau_{S})$, obtained numerically.
\end{itemize} The $D_{\alpha}(\overline{\rho}_{r}||\tau_{S})$ are the most likely candidates for observable quantities, whereas the other two assume that single-shot measurements of free energy are possible.

\subsection{Qubit Systems}
\label{sec:qubit}

We start with $d=2$. Before discussing the behavior of the free energies under perturbation, let us have a look at the thermal states. The reference state $\tau_S$ has a ground-state occupation $q_0=(1+e^{-\beta\hbar g_0})^{-1}$. The ensemble-averaged thermal state $\overline{\tau}$ has
\ba
\overline{q}_0&=&q_0\int G(\delta)\frac{1+e^{-\beta\hbar g_0}}{1+e^{-\beta\hbar g_0(1+\delta)}}\,d\delta\,.
\ea For $\beta>0$, it's easy to prove that $\overline{q}_0\leq q_0$ \footnote{The simplest is to notice that $\int G(\delta)f(\delta)\,d\delta=\int G(\delta)f_{\textrm{even}}(\delta)\,d\delta$ with $f_{\textrm{even}}(\delta)=\frac{1}{2}(f(\delta)+f(-\delta))$. For our case, with $e^{-\beta\hbar g_0}\leq 1$, it is elementary to prove that the maximum of $f_{\textrm{even}}(\delta)$ is $f_{\textrm{even}}(0)=1$, then the function decreases monotonically towards $\lim_{\delta\rightarrow\pm\infty}f_{\textrm{even}}(\delta)=\frac{1}{2}(1+e^{-\beta\hbar g_0})$.}: in other words, $\overline{\tau}$ is more mixed that $\tau_S$.

\begin{figure}[h]
\centering
\subfloat[ (Colour online) Behaviour of $D_{\infty}$ after each successive interaction
for the single-shot scenario $\rho_{r}$  (solid, black) and the analytical ensemble-averaged expression $\overline{\rho}_{r}$ (solid, green) as well as $\overline{D}_{\infty}$ (dotted, red) over 1000 runs.\label{fig:Finf_Steps_Qubits}]{\pic{trim={3ex 0 3ex 0},clip,width=0.48}{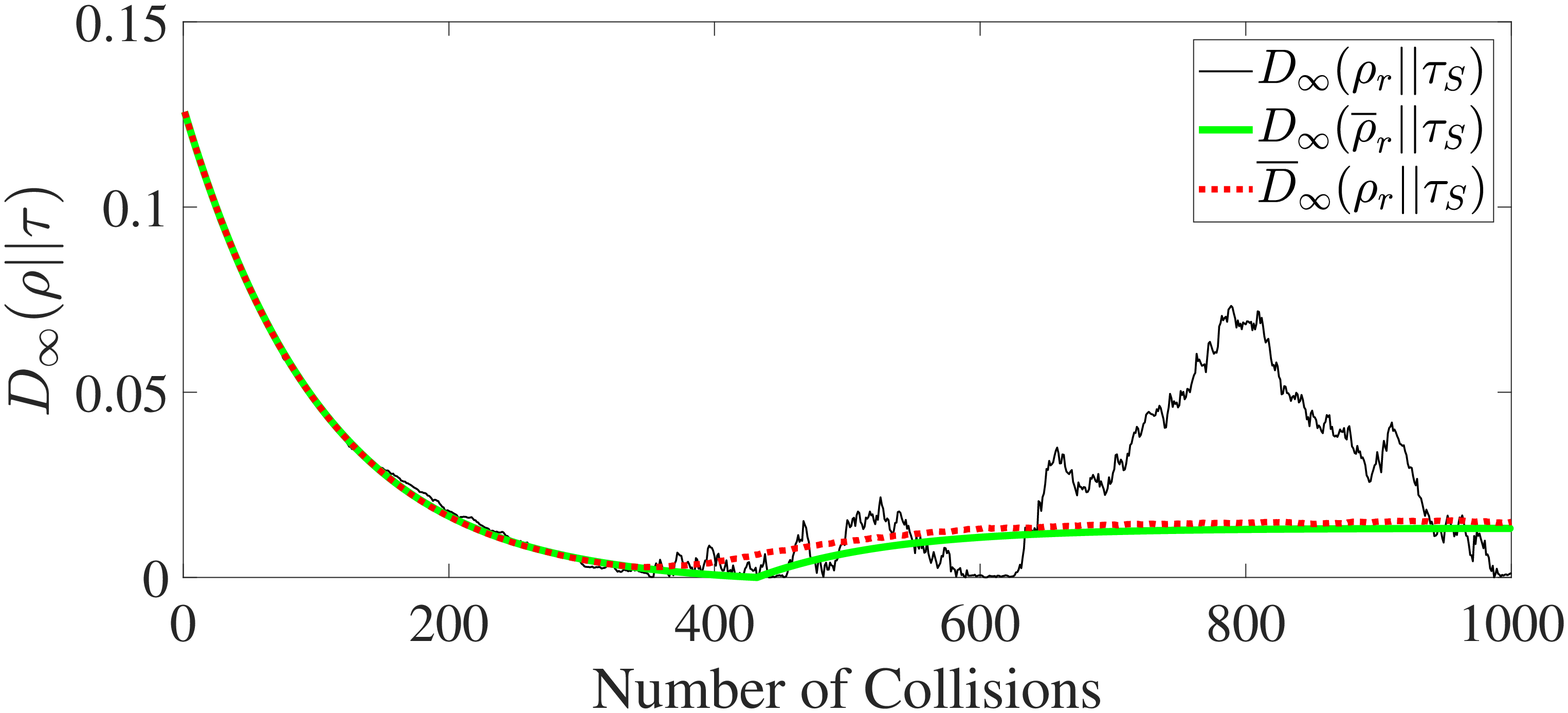}}\hfill
\subfloat[ (Colour online) Parametric plots of of $D_{1}$ (solid, red) and $D_{\infty}$ (dash-dotted, blue) with the trajectory of $\overline{\rho}_{r}$ (black dots). \label{fig:Finf_p0_Qubits}]{\pic{trim={3ex 0 3ex 0},clip,width=0.48}{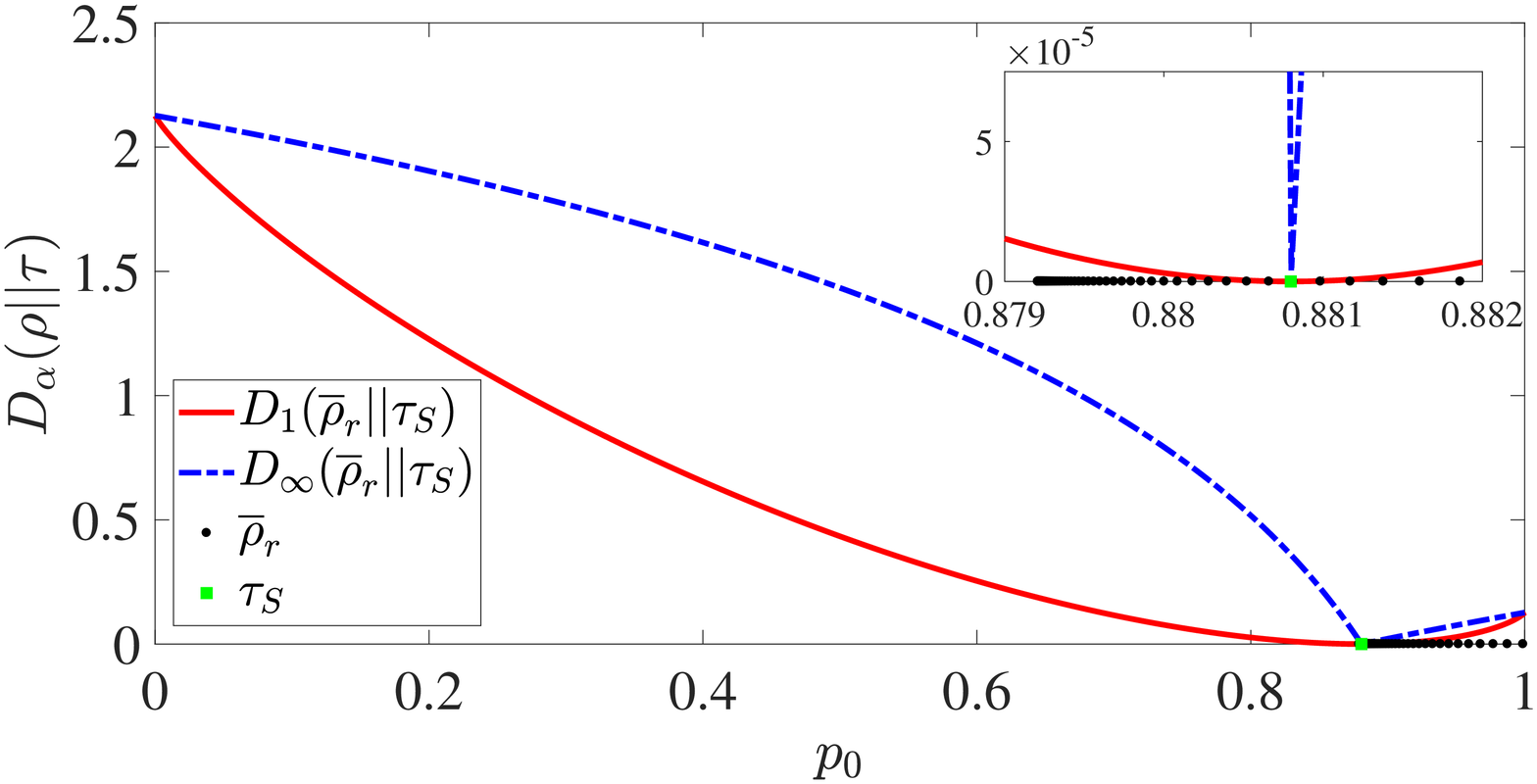}}
\caption{In these plots, $\hbar g_0=2$, $\Delta^2=0.1$, $\beta=1$, $\theta=0.1$ and $\rho_{0}$ is the ground state.}
\end{figure}

Let us now look at the dynamics \eqref{dynamics} of a state diagonal in the eigenbasis of $H_S$. In Fig.~\ref{fig:Finf_Steps_Qubits}, we plot $D_{\infty}(\rho_{r}||\tau_{S})$ for the ground state as the initial input state and for some choice of the parameters. We see an initial smooth decrease, then a partial increase which, for the case of single-shot realisations, is accompanied by fluctuations. By inspection, we have found that the behavior is qualitatively the same across parameter space, with both the amount of final increase and the size of the fluctuations being larger for larger values of $\alpha$ and for larger values of $\beta$ (low temperature).


In order to understand this behavior, we plot $D_{\alpha}(\rho_{r}||\tau_{S})$ as a function of $p_0$, which is the only free parameter for states that are diagonal in the eigenbasis of $H_S$ (Fig.~\ref{fig:Finf_p0_Qubits}). From this plot, most of the previously observed features become very transparent: we have set the evolution to start with $p_0=1$ (ground state) and we know that it must end at $p_0=\overline{q}_0<q_0$. It must therefore pass through $p_0=q_0$, where all the $D_{\alpha}$ are equal to zero. Thus, for every $\alpha>0$ and for every $\Delta^2>0$ (that is, however small the perturbation), $F_{\alpha}(\overline{\rho}_{r}||\tau_{S})$ will decrease monotonically if and only if the initial state has $p_0\leq q_0$. In other words, all the second laws for the $F_{\alpha}(\overline{\rho}_{r}||\tau_{S})$ can be violated in our model, and this is seen by taking any initial state with $p_0>q_0$ \footnote{Recall that we have assumed $\beta>0$: if one were willing to consider negative temperatures, $\overline{q}_0\geq q_0$ would hold and the conditions would be reversed.}.

We are left to explain the fluctuations of the single-shot trajectory obtained for a configuration of the $\delta_r$ chosen at random. We see in Fig.~\ref{fig:Finf_p0_Qubits} that the gradient of $F_\infty$ is very steep for $p_0<q_0$, in the vicinity of $q_0$, and less steep elsewhere. Thus, the fluctuations become significant only after the state crosses $\tau_S$. For our choice of parameters, the target state $\overline{\tau}$ is very close to $\tau_S$, and this is why the fluctuations do not subside. The generally low gradient of $F_1$ in the region of our trajectory, also shown in Fig.~\ref{fig:Finf_p0_Qubits}, explains why the fluctuations would be less prominent for that free energy.

\subsection{Qutrit Systems}
\label{sec:qutrit}

We repeat a similar study for qutrits. Because with qutrits one can do all that can be done with qubits, the main conclusion will be the same: there exist initial states such that all the second laws, except for $\alpha=0$, are violated by an arbitrary small amount of perturbation. The picture however becomes richer and is worth presenting.


\begin{figure}[ht]
\centering
\subfloat[(Colour online) Behaviour of $D_{\infty}$ after each successive interaction
for the single-shot scenario $\rho_{r}$  (solid, black) and the analytical ensemble-averaged expression $\overline{\rho}_{r}$ (solid, green) as well as $\overline{D}_{\infty}$ (dotted, red) over 1000 runs.
\label{fig:Finf_Steps_Qutrits}]{\pic{trim={3ex 0 3ex 0},clip,width=0.48}{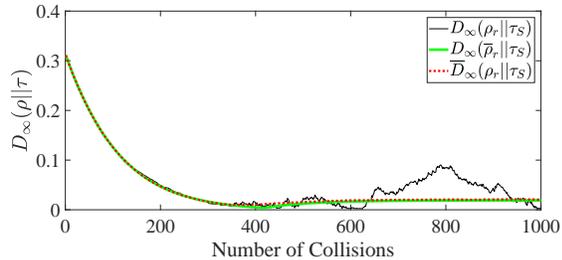}}\hfill
\subfloat[(Colour online) Contour plots for $F_{1}$ (solid, magenta) and $F_{\infty}$ (dash-dotted, black) overlaid with a single-shot trajectory $\rho_{r}$ (dark green dots) and the analytical ensemble averaged trajectory of $\overline{\rho}_{r}$ (solid, red) for ground state initial inputs. The solid light green line is a sample trajectory (with a different input state) that violates the generalized second laws for all $\alpha>0$ except $\alpha=\infty$.
\label{fig:Contour_Qutrits}]{\pic{trim={6.5cm 0cm 9.5cm 6.2cm},clip,width=0.48}{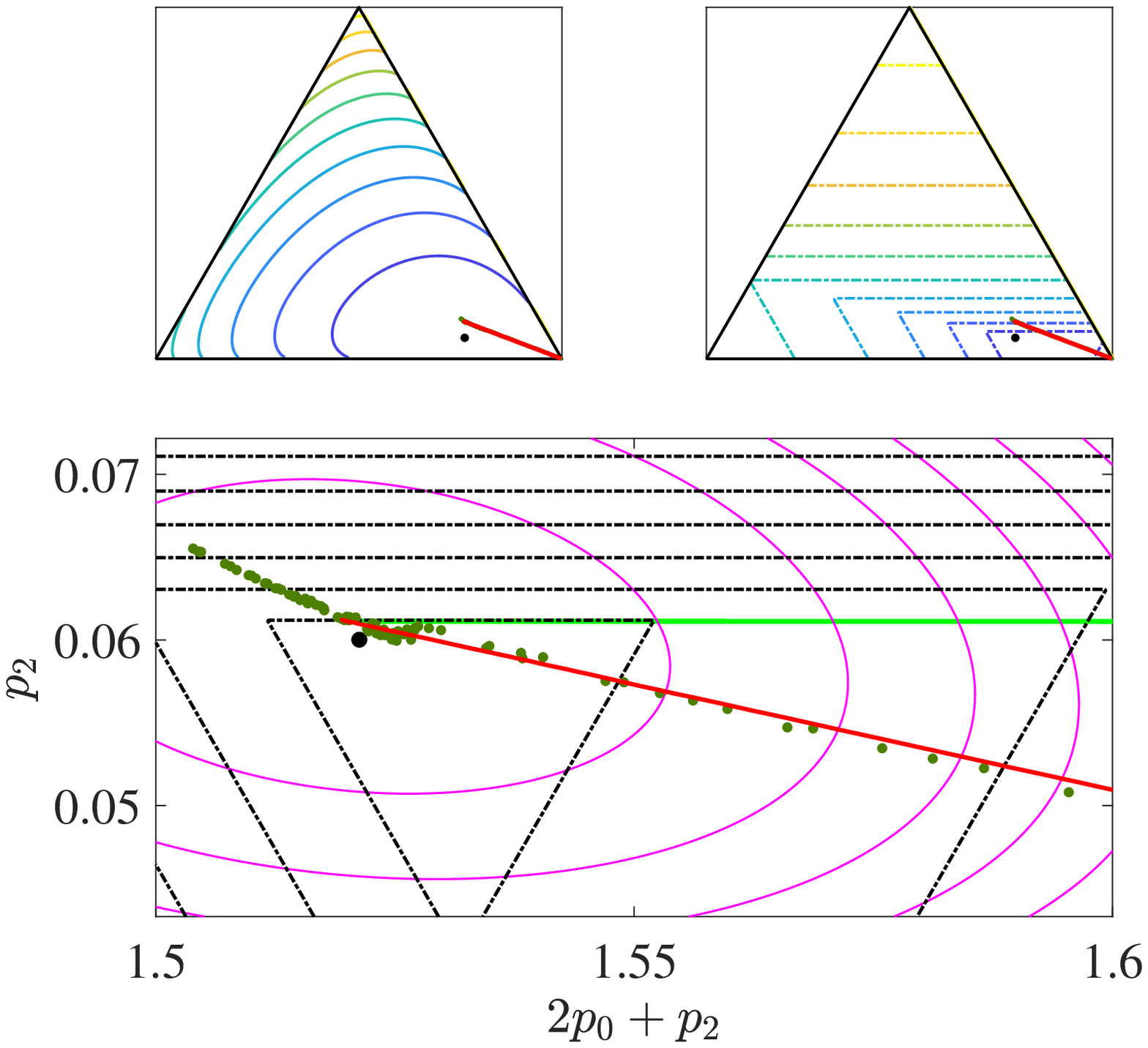}}
\caption{In these plots, $\hbar g_0=2$, $\Delta^2=0.1$, $\beta=1$ and $\theta=0.1$.}
\end{figure}

The behavior of the trajectories given the ground state as the initial state (Fig.~\ref{fig:Finf_Steps_Qutrits}) is the one we are familiar with. Like for qubits, it is very clarifying to plot the trajectories and the contour lines of the $F_\alpha$ in the parameter space. For diagonal states of qutrits, the parameter space is two-dimensional. We choose a parametric plot described in detail in Appendix \ref{apptriangles}.

For the partial swap interaction, the time evolution of ensemble-averaged state \eqref{eq:Analytic_Noise} traces a straight line in the parametric plot, between the initial state and $\overline{\tau}$. Each single-shot realization corresponds to a noisy trajectory around that line. An increase of $F_{\alpha}$ will be observed whenever the trajectory crosses some contour lines twice. This is the case for the trajectories plotted in Fig.~\ref{fig:Finf_Steps_Qutrits}, as shown in the parametric plot of Fig.~\ref{fig:Contour_Qutrits}.

Like for the case of qubits, all the second laws for $\alpha>0$ are violated if $\tau_S$ lies on the line connecting the initial and the final state. However, other trajectories may be such that only some of the second laws are violated. For instance, by taking an initial state that has the same $p_2$ as $\overline{\tau}$, the linear trajectory follows one of the contour lines of $F_{\infty}$ while violating all the second laws for finite $\alpha$ (solid light green line in Fig.~\ref{fig:Contour_Qutrits}). This example is peculiar:  on a generic trajectory, the second law associated to $F_{\infty}$ seems to be the easiest to violate (larger increase, larger fluctuations), while here it is the only one that is respected. It must also be kept in mind that the linear trajectories are proper to the choice \eqref{partialswap} of $U_{S,r}$: modifications of the collisional model would lead to curved trajectories in the parametric plot (Appendix \ref{appothers}).

\section{Physical relevance of the model}
\label{sec:PhyRel}

The collisional model that we studied is unquestionably a toy model, so it should not be over-interpreted. In particular, in a full treatment, the unitary $U$ should derive from an interaction Hamiltonian. That being said, it can be related to commonplace physical situations, and the differences do not seem to alter the heart of the matter.

A first possible physical reading is that the qubits are spins immersed in an external magnetic field $B$ pointing in the $z$ direction. In this case, the coupling is given by $g=\gamma B$ where $\gamma$ is the gyromagnetic factor. The fact that $g$ depends on $r$ can be attributed to spatial inhomogeneity of the field: in this case, however, the interaction should not be by contact, because if the two qudits are at the same position they should feel the same field. It could be attributed to inhomogenities in $\gamma$ due to different chemical environments.

A second reading is possible: since the perturbation is scalar, it appears only in products $\beta g_0 (1+\delta)$ and could therefore be attributed to $\beta$ rather than to the coupling. The auxiliary qudits are thermalised, so they have been in contact with a thermal bath. Attributing the inhomogeneity to $\beta$ amounts to saying that temperature in that bath is not homogeneous. In this reading, $[H,U]=0$ but the $\tau_r$ are thermal states at different temperatures. Thus, all the second laws can also be violated by departing from the free resources in this way.



\section{Conclusion}
\label{sec:conclusion}

We have studied the robustness of the second laws of thermal operations under the relaxation of the framework. The study is based on a toy collisional dynamics, and the deviation from the framework is in the form of inhomogeneous scalar parameters that induces $[H,U]\neq 0$.

We observe that all $\alpha>0$ second laws are violated for arbitrarily low perturbations. The violation is visible in single-shot trajectories, in the free energy of the ensemble-averaged state, and in the ensemble-averaged values of the free energy. A representation in parametric plots provides a compelling picture of the origin of this behavior.

While the violation is usually more prominent for $F_{\infty}$, it is also present for $F_1$ that is supposed to capture the ``normal version'' of the second law in the resource theory framework. There is nothing paradoxical in violating theorems by departing from their assumptions. Nonetheless, in the context of our model, the departure amounts to adding a small inhomogeneity in rather simple external environments: the phenomenological second law has acquitted itself quite well describing far more uncontrolled situations like wet environments (see \cite{Wang_2013} for a recent result).

These results indicate that the second laws of thermal operations cannot be used in the same way as the usual second law of thermodynamics. They add to the challenges posed by extending the resource theory of thermodynamics to imperfect reservoirs \cite{Sparaciari_2017}.


\section*{acknowledgments}

We acknowledge illuminating discussions with and useful feedback from Philippe Faist, Rodrigo Gallego, Kavan Modi, Markus M\"uller, Nelly Ng and Henrik Wilming.

This research is supported by the Singapore Ministry of Education Academic Research Fund Tier 3 (Grant No. MOE2012-T3-1-009); by the National Research Fund and the Ministry of Education, Singapore, under the Research Centres of Excellence programme; and by the John Templeton Foundation Grant 60607 ``Many-box locality as a physical principle''.

\begin{appendix}

\section{Further observations for qutrits (and higher dimensions)}

\subsection{The geometry of the $F_{\alpha}$}
\label{apptriangles}

\begin{center}
\begin{figure}[H]
    \pic{trim={3.6cm 0 3.5cm 0},clip,width=1}{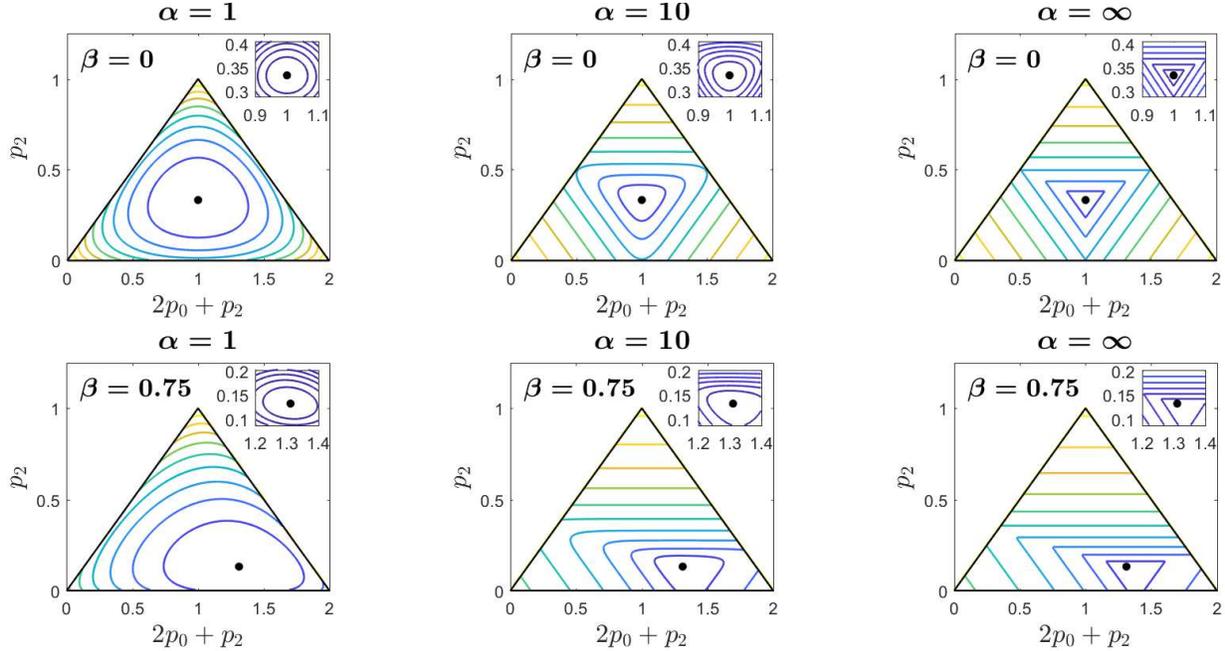}
\caption{(Colour online) Contour plots of $F_{\alpha}$ for $\beta=0$ (infinite temperature, top) and $\beta=0.75$ (Bottom). The apex of the triangle stands for $p_{2}=1$, the bottom left corner for $p_{1}=1$ and bottom right corner for $p_{0}=1$. The black dot denotes thermal state of the system. The inset shows a magnified view of the contour lines closer to the thermal state}
\label{fig:Contour_Sweep}
\end{figure}
\end{center}

Diagonal states of qutrits are parametrised by $p_0$, $p_1$ and $p_2=1-p_0-p_1$. In order to illustrate the contour lines of the $F_{\alpha}$ (the same as those of the $D_\alpha$), we choose a representation in which the three eigenstates are at the cusps of an equilateral triangle: the ground state is at the bottom right corner, the intermediate state is at the top, and the highest energy state is at the bottom left corner (Fig.~\ref{fig:Contour_Sweep}).

The minimum is always achieved for the thermal state $\tau_S$, represented by a black dot in the figure. Any free evolution will be represented by a trajectory that crosses each contour line only once, for all values of $\alpha$. The trajectories need not be straight lines, as explained in the next subsection.

\onecolumngrid
\twocolumngrid

\subsection{Modifications of the collisional model}
\label{appothers}

Throughout the paper, we have assumed all the two-body interactions $U_{S,r}$ are given by partial swap with mixing angle $\theta$, which are but a subset of the larger and more general set of energy conserving operations. The most general energy preserving unitaries are those that do not cause any mixing between the energy subspaces. This is just a direct sum of the unitaries acting on each energy subspace 
\begin{align}
    U=\bigoplus_{k=0}^{2d-2} U_k
\end{align}
where $U_{k}$ is a unitary on the span$\{\ket{jj'}\}$ such that $j+j'=k$ and $d$ is the dimension of the system. For instance, $U_{1}$ is just a generalized unitary on the subspace $\{\ket{01},\ket{10}\}$. One can determine the number of free parameters for each $U_k$ by squaring the dimension of that energy subspace. Therefore for energy subspaces of one dimension, the only free parameter will be a phase factor; for energy subspaces of two and three dimensions, any unitary with four and nine free parameters respectively will suffice. Hence in the case of qubits, there will be a total of six free parameters (five, if we exclude a global phase), and for qutrits, there would be 18 (17) free parameters.

\begin{center}
\begin{figure}[ht]
\pic{trim={8.5cm 0 8.5cm 0},clip,width=0.5}{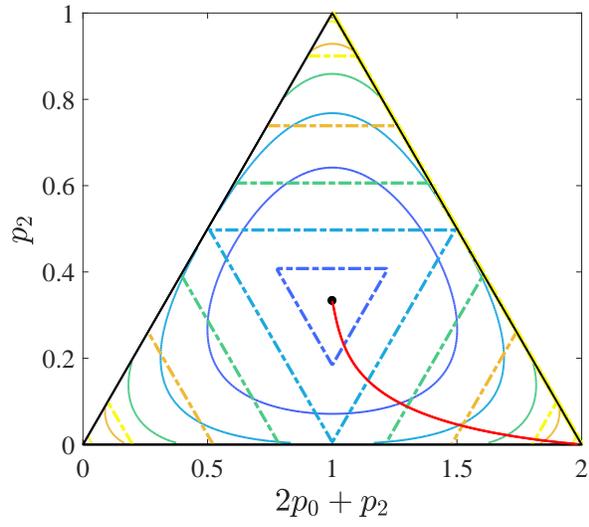}
\caption{(Colour online) Contour plots for $F_{1}$ (solid) and $F_{\infty}$ (dash-dotted) overlaid with the  ensemble averaged trajectory of $\overline{\rho}_{r}$ (solid, red) for the interaction with $\beta=0$, $hg_0=2$, $\Delta^2=0$, $\theta_{1}=0.075$, $\theta_{2}=0.05$ and $\theta_{3}=0.1$.}
\label{fig:Curved}
\end{figure}
\end{center}

Under these more general unitaries, the free evolution does not need to be a straight line as defined by Eq.~\eqref{eq:Analytic} for the partial swap \eqref{partialswap}. For instance, with partial swaps of different mixing angles $\theta_{k}$ between each energy subspace, one can obtain curved trajectories as in Fig.~\ref{fig:Curved}.

\end{appendix}


\begin{thebibliography}{3}%
\makeatletter
\providecommand \@ifxundefined [1]{%
 \@ifx{#1\undefined}
}%
\providecommand \@ifnum [1]{%
 \ifnum #1\expandafter \@firstoftwo
 \else \expandafter \@secondoftwo
 \fi
}%
\providecommand \@ifx [1]{%
 \ifx #1\expandafter \@firstoftwo
 \else \expandafter \@secondoftwo
 \fi
}%
\providecommand \natexlab [1]{#1}%
\providecommand \enquote  [1]{``#1''}%
\providecommand \bibnamefont  [1]{#1}%
\providecommand \bibfnamefont [1]{#1}%
\providecommand \citenamefont [1]{#1}%
\providecommand \href@noop [0]{\@secondoftwo}%
\providecommand \href [0]{\begingroup \@sanitize@url \@href}%
\providecommand \@href[1]{\@@startlink{#1}\@@href}%
\providecommand \@@href[1]{\endgroup#1\@@endlink}%
\providecommand \@sanitize@url [0]{\catcode `\\12\catcode `\$12\catcode
  `\&12\catcode `\#12\catcode `\^12\catcode `\_12\catcode `\%12\relax}%
\providecommand \@@startlink[1]{}%
\providecommand \@@endlink[0]{}%
\providecommand \url  [0]{\begingroup\@sanitize@url \@url }%
\providecommand \@url [1]{\endgroup\@href {#1}{\urlprefix }}%
\providecommand \urlprefix  [0]{URL }%
\providecommand \Eprint [0]{\href }%
\providecommand \doibase [0]{http://dx.doi.org/}%
\providecommand \selectlanguage [0]{\@gobble}%
\providecommand \bibinfo  [0]{\@secondoftwo}%
\providecommand \bibfield  [0]{\@secondoftwo}%
\providecommand \translation [1]{[#1]}%
\providecommand \BibitemOpen [0]{}%
\providecommand \bibitemStop [0]{}%
\providecommand \bibitemNoStop [0]{.\EOS\space}%
\providecommand \EOS [0]{\spacefactor3000\relax}%
\providecommand \BibitemShut  [1]{\csname bibitem#1\endcsname}%
\let\auto@bib@innerbib\@empty
\bibitem [{Note1()}]{Note1}%
  \BibitemOpen
  \bibinfo {note} {We notice that, in a dynamical system, the approach to
  energy conservation is different. There, the whole evolution is generated by
  the Hamiltonian that includes interactions. Energy is then conserved if the
  total Hamiltonian is time-independent; if it is time-dependent, there is no
  reason for energy to be conserved.}\BibitemShut {Stop}%
\bibitem [{Note2()}]{Note2}%
  \BibitemOpen
  \bibinfo {note} {The simplest is to notice that $\DOTSI \intop \ilimits@
  G(\delta )f(\delta )\protect \tmspace +\thinmuskip {.1667em}d\delta =\DOTSI
  \intop \ilimits@ G(\delta )f_{\protect \textrm {even}}(\delta )\protect
  \tmspace +\thinmuskip {.1667em}d\delta $ with $f_{\protect \textrm
  {even}}(\delta )=\protect \frac {1}{2}(f(\delta )+f(-\delta ))$. For our
  case, with $e^{-\beta \hbar g_0}\leq 1$, it is elementary to prove that the
  maximum of $f_{\protect \textrm {even}}(\delta )$ is $f_{\protect \textrm
  {even}}(0)=1$, then the function decreases monotonically towards $\protect
  \qopname \relax m{lim}_{\delta \rightarrow \pm \infty }f_{\protect \textrm
  {even}}(\delta )=\protect \frac {1}{2}(1+e^{-\beta \hbar g_0})$.}\BibitemShut
  {Stop}%
\bibitem [{Note3()}]{Note3}%
  \BibitemOpen
  \bibinfo {note} {Recall that we have assumed $\beta >0$: if one were willing
  to consider negative temperatures, $\protect \overline {q}_0\geq q_0$ would
  hold and the conditions would be reversed.}\BibitemShut {Stop}%
\end{thebibliography}%


\begin{thebibliography}{00}

\bibitem{Brandao_2013} 
F.G.S.L.~Brand\~{a}o, M.~Horodecki, J.~Oppenheim, J.M.~Renes, R.W.~Spekkens, \textit{The Resource Theory of Quantum States Out of Thermal Equilibrium}, Phys. Rev. Lett. \textbf{111}, 250404 (2013)

\bibitem{Gour_2015}
G.~Gour, M.P.~Müller, V.~Narasimhachar, R.W.~Spekkens, N.~Yunger Halpern, \textit{The resource theory of informational nonequilibrium in thermodynamics}, Phys. Rep. \textbf{583}, 1-58 (2015) 

\bibitem{Goold_2016}
J.~Goold, M.~Huber, A.~Riera, L.~del Rio, P.~Skrzypczyk, \textit{The role of quantum information in thermodynamics --- a topical review}, J. Phys. A: Math. Theor. \textbf{49}, 143001 (2016)

\bibitem{Sparaciari_2017}
C.~Sparaciari, D.~Jennings and J.~Oppenheim \textit{Energetic instability of passive states in thermodynamics}, Nature Comm. \textbf{8}, 1895 (2017)

\bibitem{Meer_2017}
R van der Meer, N.~Ng and S.~Wehner, \textit{Smoothed generalized free energies for thermodynamics}, Phys. Rev. A \textbf{96}, 062135 (2017).

\bibitem{Mueller_2017}
M.~P.~M\"{u}ller, \textit{Correlating Thermal Machines and the Second Law at the Nanoscale}, \href{https://arxiv.org/abs/1707.03451}{arXiv:1707.03451}

\bibitem{Baumer_2017}
E.~B\"{a}umer, M.~Perarnau-Llobet, P.~Kammerlander and R. Renner, \textit{Partial Thermalizations Allow for Optimal Thermodynamic Processes} \href{https://arxiv.org/abs/1712.07128}{arXiv:1712.07128}

\bibitem{Brandao_2015}
F.~Brand\~{a}o, M. Horodecki, N. Ng, J. Oppenheim, and S. Wehner, \textit{The second laws of quantum thermodynamics}, Proceedings of the National Academy of Sciences, \textbf{112}, 3275-3279 (2015)

\bibitem{Skrzypczyk_2014}
P.~Skrzypczyk, A.~J.~Short and S.~Popescu, \textit{Work extraction and thermodynamics for individual quantum systems}, Nature Comm. \textbf{5}, 5185 (2014)


\bibitem{Valerio_2002}
V.~Scarani, M.~Ziman, P.~\v{S}telmachovi\v{c}, N. Gisin and V. Bu\v{z}ek, \textit{Thermalizing Quantum Machines: Dissipation and Entanglement} Phys. Rev. Lett. \textbf{88}, 097905 (2002).

\bibitem{Wang_2013} Z.~Wang, R.~Hou, A.~Efremov, \textit{Directional fidelity of nanoscale motors and particles is limited
by the 2nd law of thermodynamics—Via a universal equality}, J. Chem. Phys. \textbf{139}, 035105 (2013)




\end{thebibliography}
\end{document}